\documentclass{article}
\usepackage{authblk}
\usepackage[english]{babel}
\usepackage[letterpaper]{geometry}
\usepackage{amsmath,amssymb}
\usepackage{graphicx}
\usepackage[colorlinks=true, allcolors=blue]{hyperref}

\DeclareMathOperator{\real}{Re}

\title{Precision Electroweak Physics of Neutral Currents Below the Z-pole}

\author[1]{Jens Erler} 
\author[2]{K. S. Kumar} 
\affil[1]{PRISMA$^+$ Cluster of Excellence, 
Institute for Nuclear Physics and Helmholtz-Institut Mainz, \newline
Johannes Gutenberg-University, 55099 Mainz, Germany} 
\affil[2]{Amherst Center for Fundamental Interactions and Physics Department, \newline
University of Massachusetts, Amherst, MA 01003, USA}

\begin{document}
\maketitle
\begin{abstract}
We report on the status and prospects of precision weak neutral current interaction 
measurements in fixed target scattering, atomic parity violation, as well as future collider 
measurements at momentum transfer $\sqrt{Q^2} \ll M_Z$, including the experiments 
in parity-violating electron scattering under construction at Jefferson Laboratory and 
at Mainz that are expected to commence data collection in the very near future.
\end{abstract}

\section{Introduction} 
With currently only a few interesting deviations of modest statistical significance, 
the Standard Model (SM) of leptons~\cite{Weinberg:1967tq} and quarks~\cite{Glashow:1970gm} has 
successfully passed all experimental tests so far.
Moreover, no new particles turned up at the LHC.  
On the other hand, the SM must be considered incomplete, as it does not provide an explanation 
for the ever increasing evidence for dark matter~\cite{Bertone:2016nfn}.
Thus, the SM well approximates current laboratory experiments, but at some level small 
deviations are expected to show up in high precision measurements when compared to SM 
calculations.
Such precision measurements are possible at low momentum transfer provided that the usually 
strongly dominant electromagnetic (QED) and strong (QCD) interaction effects can be screened.  
This is possible by benefiting from the parity-violating nature of the weak interaction in 
polarized electron scattering (Section~\ref{PVES}) and atomic physics (Section~\ref{APV}), 
or by utilizing neutrino projectiles (Section~\ref{neutrinos}).

Of particular interest are neutral current interactions~\cite{Glashow:1961tr}, 
\begin{equation}
\label{NCLagrangian}
{\mathcal L}_{\rm NC} = - e \sum_i Q_i\, \overline{\psi}_i\, \gamma^\mu\, \psi_i\, A_\mu - 
\frac{g}{2\cos\theta_W} \sum_i \overline{\psi}_i\, \gamma^\mu (g^i_V - g^i_A \gamma^5 )\, 
\psi_i\, Z_\mu
\end{equation}
mediated at tree level by photon ($A_\mu$) and $Z$ boson exchange, 
\begin{equation}
\label{gaugefields} 
	A_\mu \equiv B_\mu \cos\theta_W + W_\mu^3 \sin\theta_W \hskip 60pt
	Z_\mu \equiv - B_\mu \sin\theta_W + W_\mu^3 \cos\theta_W
\end{equation}
where $\theta_W$ is the weak mixing angle, 
$e$ ($g$) is the electromagnetic (weak isospin) gauge coupling, 
and $g_V^i$ and $g_A^i$ are the vector and axial-vector $Z$ couplings to fermion $\psi_i$.
$B_\mu$ and $W_\mu^3$ are the gauge bosons corresponding to $U(1)$ hypercharge and the third 
component of weak isospin, respectively.

In discussing electroweak precision observables at low ($Q^2 \ll M_Z^2$) momentum 
transfers~\cite{Erler:2004cx} (for precision measurements of $\theta_W$ in the $Z$ pole region, 
see the contribution by Arie Bodek, Hyon-San Seo, and Un-Ki Yang in this volume), it is 
convenient to introduce four-fermion interactions relevant to $\nu$-hadron, $\nu e$, as well as 
parity-violating $e$-hadron and $ee$ neutral-current processes in the form~\cite{Erler:2013xha},
\begin{align}
{\cal L}^{\nu e } =& - \frac{1}{2 v^2}\ \bar\nu\, \gamma_\mu (1 - \gamma^5 ) \nu\; \bar e\, 
\gamma^\mu ( g^{\nu e}_{LV} - g^{\nu e}_{LA}\, \gamma^5 )e \\
{\cal L}^{\nu h} =& - \frac{1}{2 v^2}\ \bar\nu\, \gamma_\mu (1 - \gamma^5 ) \nu \sum_q 
\left[ g_{LL}^{\nu q}\, \bar q\, \gamma^\mu (1 - \gamma^5 ) q + 
g_{LR}^{\nu q}\, \bar q\, \gamma^\mu (1 + \gamma^5 ) q \right] \\ 
{\cal L}^{ee} =& \frac{1}{2 v^2}\ g_{AV}^{ee}\, \bar e\, \gamma_\mu \gamma^5 e\; \bar e\, 
\gamma^\mu e \\
{\cal L}^{eh} =& \frac{1}{2 v^2} \sum_q 
\left[ g_{AV}^{eq}\, \bar e\, \gamma_\mu \gamma^5 e\; \bar q\, \gamma^\mu q + 
g_{VA}^{eq}\, \bar e\, \gamma_\mu e\; \bar q\, \gamma^\mu \gamma^5 q \right]
\end{align}
where $v \approx 246$~GeV is the Higgs vacuum expectation value, and where one must include 
the charged-current contribution for $\nu_e e$ and the parity-conserving QED contribution for 
electron scattering.
The four-fermion couplings~\cite{Erler:2013xha} appearing above are defined in 
the static limit ($Q^2 \to 0$) with specific radiative corrections included, 
while others (more experiment specific ones) are assumed to be removed by the experimentalists.
They are convenient in that their determinations from very different types of processes such as
those discussed below can be straightforwardly combined.
Note that they differ from the respective products of the respective $g_V^i$ and $g_A^i$ in 
Eq.~\eqref{NCLagrangian} in the radiative corrections and in the presence of possible physics 
beyond the SM (BSM).

\begin{figure}[t]
\label{Rodolfo}
\centering
\includegraphics[width=0.88\linewidth]{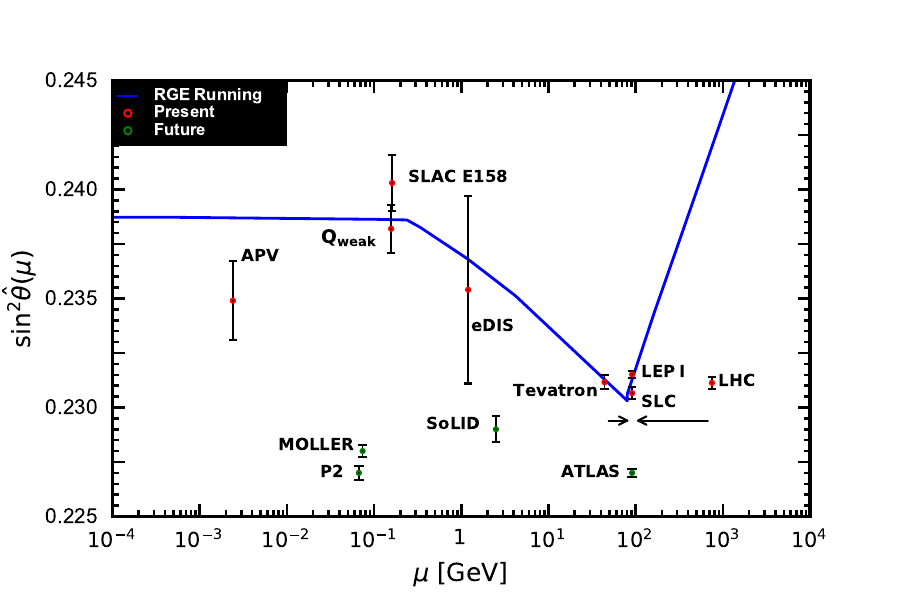}
\caption{Renormalization scale dependence (running) of $\sin^2\hat\theta_W(\mu)$ and various 
anticipated (shown at arbitrary vertical positions) and current experimental 
determinations~\cite{Erler:2017knj}.
The hadron collider results from the Tevatron and the LHC are from Drell-Yan production of 
charged lepton pairs in an invariant mass window around the $Z$ mass, $M_Z$, 
but shifted to the left and right, respectively, to avoid clutter.
The nominal uncertainty in the predicted running is below the thickness of the blue curve.
However, the very recent lattice gauge theory calculation~\cite{Boccaletti:2024guq} of the 
hadronic vacuum polarization contribution to the anomalous magnetic moment is in disagreement 
with the traditional evaluation of a dispersion integral based on cross section measurements of 
$e^+ e^- \to$ hadrons (as well as hadronic $\tau$ decay spectral functions).
This issue enters in a correlated manner into the calculation of 
$\sin\hat\theta_W(0)$.  The difference in the calculated values of 
$\sin\hat\theta_W(0)$ is about $7 \times 10^{-5}$~\cite{Erler:2024lds}.}
\end{figure}

If in some low energy semi-leptonic process the $Z$ boson couples through the $g_{AV}^{eq}$ 
(but not the $g_{VA}^{eq}$) then some of the non-renormalization theorems 
(historically referred to as the conserved vector current hypothesis~\cite{Gershtein:1955fb}) 
carry over from QED.
Furthermore, the effective electron and proton couplings, $g_{AV}^{ee}$ and $g_{AV}^{ep}$,
are proportional to $1 - 4 \sin^2\hat\theta_W(\mu)$ at tree level, where $\hat\theta_W(\mu)$ 
denotes the weak mixing angle in the $\overline{\rm MS}$ renormalization scheme.
It depends on the t'Hooft scale $\mu$, which should be chosen close to the typical momentum 
transfer of the process, so as to avoid spurious logarithms spoiling the accuracy of 
the SM calculations.
Figure~\ref{Rodolfo} shows that $\sin^2\hat\theta_W(\mu)$ is numerically quite close to 1/4,
implying an enhanced sensitivity not only to $\hat\theta_W(\mu)$, 
but also to the possible presence of physics beyond the SM.
The figure also shows that this enhancement is even more pronounced at the low scales relevant
for the MOLLER and P2 experiments discussed in Section \ref{PVES}.

The $1 - 4 \sin^2\hat\theta_W(\mu)$ suppression in $g_{AV}^{ee}$ and $g_{AV}^{ep}$ also 
translates to a relative enhancement of BSM sensitivity.  
Indeed, while in collider experiments with $Q^2 \sim M_Z^2$ new physics contributions to 
four-fermion amplitudes are parametrically suppressed by $\Gamma_Z^2/M_Z^2$, at low energies 
they are only power-suppressed compared to the SM amplitudes.
BSM physics can typically be tested up to scales of order $v/\sqrt{\sigma}$ (or higher in case 
of strongly coupled new physics or of the $1 - 4 \sin^2\hat\theta_W(\mu)$ suppression), where 
$v \approx 246$~GeV is the Higgs vacuum expectation value and $\sigma$ is 
the {\em relative\/} experimental uncertainty.

\section{Parity-Violating Electron Scattering (PVES)}
\label{PVES}
\subsection{Overview}
The technique of electron scattering has been used for more than sixty years to study nuclear 
and nucleon structure using the experimental cleanliness and well-understood theory of 
electromagnetic interactions. 
Soon after parity-violation in the weak interactions was demonstrated, Zel'dovich noted that 
if an analogous neutral interaction existed, then parity violation would be manifested in 
longitudinally polarized electron scattering off unpolarized nucleons due to the interference 
between the weak and electromagnetic amplitudes.
He estimated that the parity-violating asymmetry $A_{PV}$, {\em i.e.}, the fractional difference 
in the cross-sections when the longitudinal polarization of the electron beam is reversed, 
would be of order $10^{-4}\, Q^2$ GeV$^{-2}$~\cite{Zeldovich}.

For typical fixed target experiments, $A_{PV}$ ranges from roughly $10^{-4}$ 
to as small as $10^{-7}$. 
Parity violation in deep inelastic electron nucleon scattering was first observed by 
the SLAC E122 experiment~\cite{Prescott:1978tm}, constituting an important confirmation of 
the SU(2)$\times$U(1) gauge theory of electroweak interactions.
Over the past 40 years, the experimental techniques employed to measure these tiny 
left-right asymmetries to study semileptonic and purely leptonic neutral weak interactions have 
been steadily refined such that statistical and systematic errors 
better than 1 part per billion (ppb) are possible~\cite{Kumar:2013yoa,Erler:2014fqa}. 
With judiciously chosen targets and kinematics, this has facilitated measurements in several
important physics topics, such as many-body nuclear physics, nucleon structure and searches 
for physics beyond the standard model (BSM) at the TeV scale.

In the following subsections, we elaborate on the last topic, namely BSM searches. 
Weak neutral current (WNC) interactions at $Q^2\ll M_Z^2$ can probe for new four-fermion contact 
interactions generated for example by new heavy Z$^\prime$ bosons or leptoquarks whose effects 
might be highly suppressed in measurements on 
the $Z$ pole~\cite{Kumar:2013yoa,Erler:2014fqa,Ramsey-Musolf:1999qyv,Erler:2003yk}, and for 
dark sector mediators that have small admixtures to the $Z$ boson~\cite{Davoudiasl:2012ag}. 
Since the neutral current amplitude at the $Z$ pole is imaginary, 
there are no resonance interference terms with new, real four-fermion amplitudes. 
At low $Q^2$ on the other hand, interference effects might be measurable if sufficient accuracy 
is achieved.

\subsection{Past Results}
\subsubsection{SLAC E158}
While the possibility was considered of measuring parity violation in fixed target 
electron-electron (M{\o}ller) scattering soon after the SLAC E122 result, due to several factors 
that suppress the size of $A_{PV}$, the first ever measurement was carried out 
more than two decades later. 
Firstly, $A_{PV}$ in M\o ller scattering is proportional to 
$g_{AV}^{ee}G_FQ^2$~\cite{Derman:1979zc}. 
In addition to the $1 - 4 \sin^2\hat\theta_W(\mu)$ suppression in $g_{AV}^{ee}$ already 
mentioned, $Q^2$ for high energy electrons scattering off electrons of mass $m_e$ in a fixed 
target is $2m_eE_{beam}$ and therefore quite small. 
A measurement in collider mode is challenging since the increase of $Q^2$ by several orders of 
magnitude cannot overcome the large reduction in luminosity.

An experimental measurement became feasible~\cite{Kumar:1995ym} after the upgrade of the 
SLAC linac to 48~GeV by utilizing a long liquid hydrogen target to obtain sufficient luminosity; 
a successful M\o ller $A_{PV}$ measurement was carried out by 
the E158 experiment~\cite{SLACE158:2005uay}. 
Many technical developments were pioneered in the production and monitoring of the highly 
polarized electron beam, a high luminosity target, 
a novel spectrometer and detection techniques. 

The experiment was performed with 45 and 48 GeV polarized electrons, generated via 
photo\-emission of circularly polarized laser light incident on a GaAs photocathode, 
that scattered from a 1.5 m long hydrogen target. 
M\o ller-scattered electrons with energies between 13 and 24 GeV 
($4.4 < \theta_{\mathrm lab} < 7.5$ mrad) were focused onto a copper and 
quartz fiber calorimeter by a quadrupole spectrometer.  
The helicity of the beam was reversed pseudo-randomly for each beam pulse 
(120 Hz repetition rate).   
The electron beam polarization was extracted via dedicated calibration runs measuring 
the double-spin M{\o}ller scattering asymmetry from a thin magnetized foil.

Care was taken to eliminate false asymmetries due, for example, to differences in beam 
properties correlated with helicity by monitoring the position, angle, and energy with nanometer 
sensitivity; small corrections were made based on regular calibrations.  
In addition, the asymmetry was reversed every few runs by inserting a half-wave plate into 
the laser beam.  
Finally, the helicity of the beam was opposite for each of the two energies due to a $g-2$ flip 
in the magnets in the beam switchyard; hence the data were collected with roughly equal 
statistics at the two different beam energies (45 and 48 GeV).  
The asymmetry had the same magnitude and the correct sign for each of the four running 
configurations (half-wave plate state and beam energy), giving confidence in the results and 
suppressing a wide variety of possible small spurious effects.

The result of the experiment was 
\begin{equation}
A_{PV}\ (e^-e^-\rightarrow e^-e^-)=-131\pm14~{\rm (stat)}\pm10~{\rm (sys)\ ppb.}
\label{E158apv}
\end{equation} 
Care was taken to include full electroweak radiative corrections, including hard bremsstrahlung 
in the kinematic coverage~\cite{Kolomensky:2005ja}, yielding the value of 
$\sin^2\hat\theta_W(\mu=161\ \mathrm{MeV}) = 0.2403\pm 0.0013$.
One also extracts the model-independent effective coupling, $g_{AV}^{ee}= 0.0190\pm 0.0027$.
The 3\%\ shift in the running of $\sin^2\hat\theta_W$ to low $Q$ results~\cite{Czarnecki:1995fw} 
in a shift in $g_{AV}^{ee} \propto 1 - 4 \sin^2\hat\theta_W$ of about 40\%. 
Indeed, the tree-level prediction for $A_{PV}$ at the specific experimental kinematics of E158 
is about 250 ppb; the measured result demonstrated the running of $\sin^2\hat\theta_W$ 
unambiguously (more than 6$\sigma$) for the first time.

\subsubsection{JLab Qweak}
The Qweak experiment~\cite{Qweak:2014xey} was designed to measure the proton's weak charge 
$Q_W^p \approx 1 - 4 \sin^2\hat\theta_W$ via $A_{PV}$ in elastic electron-proton scattering. 
The experiment was conceived in the early 2000's to achieve 
{$\delta(A_{PV})\approx \pm 3$\% (stat.)\/}\ and $\pm 2.5$\% (syst.), was constructed between 
2006 and 2009, and data collection was completed in 2012 in two run periods lasting about 
11 months in total. 
The Qweak incident beam energy at the Thomas Jefferson National Accelerator Facility (JLab) 
was 1.165 GeV. 
Elastically scattered electrons in the range 
$\theta_{lab} = 8^\circ \pm 2^\circ\rightarrow\langle Q^2\rangle = 0.026$~GeV$^2$ were selected. 
The theoretical prediction at this $Q^2$ is $A_{PV}\approx -230$ ppb, the piece proportional to 
$Q_W^pG_F$ is about $-150$ ppb. 

In Hall C at JLab, the 1 GeV 89\%\ longitudinally polarized electron beam, with a current 
of $150-180 \mu$A, was incident on a 35 cm liquid hydrogen target. 
Elastically scattered electrons were focused by the spectrometer/collimator system onto 
an azimuthally symmetric (with respect to the beam axis) arrangement of quartz bar integrating 
Cherenkov detectors. 
The electron beam helicity was reversed pseudo-randomly in a quartet pattern at 960 Hz.
Methods similar to those developed for SLAC E158 were employed to reduce the sensitivity 
of the measured asymmetry to helicity-correlated beam fluctuations. 
Also, a new method for reversing the relative direction between the spin and momentum vectors of 
the incident electrons before acceleration (so-called Double Wien filter) was employed every few 
weeks to gain further suppression. 
In addition to using M\o ller polarimetry every few days, Qweak used a Compton polarimeter that 
monitored the electron beam polarization continuously, concomitant with physics data collection.
The absolute value of $\langle Q^2\rangle$ was calibrated in separate low current runs using 
special purpose drift-chambers that could track individual scattered electrons. 

The Qweak result~\cite{Qweak:2018tjf} was 
\begin{equation}
A_{PV}\ (e^-p\rightarrow e^-p)=-226.5\pm 7.3~{\rm (stat)}\pm 5.8~{\rm (sys)\ ppb}
\label{Qweakapv}
\end{equation} 
after all experimental corrections and electromagnetic radiative corrections were applied. 
In order to extract the proton weak charge $Q_W^p$, the small extrapolation from the finite 
$Q^2$ to zero was obtained from a fit to a range of forward angle $A_{PV}$ measurements to 
constrain the small proton structure correction with little additional uncertainty, yielding 
a measurement of $Q_W^p = 0.0719\pm 0.0045$ and $\sin^2\hat\theta_W(\mu=0) = 0.2383\pm 0.0011$.

\subsubsection{JLab PVDIS}
The JLab PVDIS~\cite{PVDIS:2014cmd, Wang:2014guo} experiment measured parity violation in 
deep-inelastic scattering off deuterium, the first such measurement since the pioneering 
SLAC E122 experiment. 
The experiment used the High Resolution Spectrometers (HRS)~\cite{Alcorn:2004sb} in Hall A 
at JLab with a 105 $\mu$A 6 GeV electron beam incident on a 20 cm long liquid deuterium target. 
The two spectrometers were set at various kinematic settings, with scattered electron momenta in 
the range 2.6 to 4 GeV and scattering angles ranging between 12.9$^\circ$ and 20.0$^\circ$. 
Since the momentum transfer $Q^2$ was of order $\sim 1$ GeV$^2$, the measured asymmetries were 
of the order of $10^{-4}$, significantly larger than previous PVES experiments that had already 
been carried out at JLab (HAPPEX3~\cite{HAPPEX:2011xlw} and PREX~\cite{Abrahamyan:2012gp}); 
the PVDIS experiment benefited from the infrastructure development for systematic control. 

The scattered electron rates ranged from 10 to 600 kHz, low enough that counting of individual 
scattered events was feasible with a dedicated electronics readout chain. 
The new challenge of the experiment was a reliable rejection of the large pion electro- and 
photo-production background while maintaining high efficiency for electron detection. 
The existing electronics had to be upgraded to count rates up to 600 kHz with hardware-based 
particle identification with minimum deadtime. 
The HRS pair each consisted of quadrupoles (Q) and a superconducting dipole (D) setup in a QQDQ 
configuration. 
Downstream of this was a pair of drift chambers to identify tracks, 
followed by a gas Cherenkov detector and a double-layered lead glass calorimeter. 

Among the measurements made at various kinematic settings, the two important kinematic points 
were in the deep-inelastic region, {\em i.e.}, where the average value of $W$, the invariant 
mass of the recoiling nucleon remnants, is greater than 2 GeV. 
Those results, shown at their respective average $Q^2$ in GeV$^2$ and Bjorken-$x$ 
(the fractional momentum carried by the struck quark), were
\begin{equation}
\begin{aligned}
A_{PV} (e^-D\rightarrow e^-X, Q^2 = 1.1, x = 0.24) = -91.1\pm 4.3 \ \rm {ppm}  \\
A_{PV} (e^-D\rightarrow e^-X, Q^2 = 1.9, x = 0.3) = -160.8\pm 7.1 \ \rm {ppm}
\end{aligned}
\label{JLabPVDIS}
\end{equation}

\subsubsection{New Physics Limits}
From these measurements, the first meaningful constraints on a linear combination of 
$g_{VA}^{eu}$ and $g_{VA}^{ed}$ were obtained. 
Comparing these results to the Standard Model expectation allows constraints on 
the corresponding four-fermion contact interaction scales $\Lambda$. 
The numerical values of these scales depend on the particular new physics model, or in the case 
of model-independent bounds on the chosen normalization.
Since in the absence of BSM signals these are {\em upper\/} bounds, 
one often uses a normalization appropriate for strongly interacting types of new physics.
For example, in a normalization~\cite{Erler:2013xha,Erler:2014fqa} in which at the tree-level,
$$ g^{ee}_{VA} = \frac{1}{2} (1 - 4 \sin^2\theta_W) $$
we normalize the new physics contact interaction according to
\begin{equation}
\label{E158+}
\Lambda_+ > v \sqrt{\frac{8\pi}{|\delta g_{VA}^{ee}|^+}} = 20.9~{\rm TeV} \qquad\qquad
\Lambda_- > v \sqrt{\frac{8\pi}{|\delta g_{VA}^{ee}|^-}} = 13.7~{\rm TeV}
\end{equation}
where
\begin{equation}
|\delta g_{VA}^{ee}|^+ = 0.0035 \qquad\qquad
|\delta g_{VA}^{ee}|^- = 0.0081
\end{equation}
are the 95\% CL upper and lower limits on non-standard contributions derived from SLAC--E158.
These limits are model-independent and only assume that the new physics is associated with
a scale $\Lambda$ much larger than the electroweak scale.
This means that rather than referring to a single new physics operator 
(of dimension six in the Standard Model Effective Field Theory),
$|\delta g_{VA}^{ee}|^\pm$ denote the limits on the linear combination of such operators 
to which the observable is most sensitive to.
Likewise, the Qweak result, $2 \delta g_{AV}^{eu} + \delta g_{AV}^{ed}= -0.0007 \pm 0.0023$ 
translates into the bounds,
\begin{equation}
\Lambda_+ > v \sqrt{\frac{\sqrt{5}\, 8\pi}{|2\, \delta g_{AV}^{eu} + \delta g_{AV}^{ed}|^+}} = 
29~{\rm TeV} \qquad\qquad 
\Lambda_- > 26~{\rm TeV}
\label{Qweak}
\end{equation}
where we used the SM prediction $Q_W^p = 0.0705$~\cite{PDG10}.
Finally, the most precisely determined PVDIS data point~\cite{PVDIS:2014cmd,Wang:2014guo} 
results in new physics constraints $\Lambda$ of ${\cal O} (8 \mbox{ TeV})$.

\subsection{Ongoing Program}
\subsubsection{MOLLER}
The MOLLER (Measurement of Lepton Lepton Electroweak Reaction) experiment~\cite{MOLLER:2014iki} is an approved and fully funded project, currently 
under construction, proposing to measure $A_{PV}$ in M\o ller scattering a factor of 5 better 
than the E158 result. 
The goal is a 2.3\%\ measurement of $Q_W^e$ resulting in 
$\delta(\sin^2\theta_W)_{\rm stat}\approx 0.00028$, which is of ${\cal{O}}(\pm0.1\%)$ and 
similar to the two best high energy collider extractions of the parameter from measurements 
at LEP and the SLC. 
The constraints \eqref{E158+} will improve considerably with the anticipated MOLLER result 
to reach $\Lambda \sim {\cal O} (40 \mbox{ TeV})$. 
This would be the most sensitive probe of new flavor- and CP-conserving neutral current 
interactions in the leptonic sector until the advent of a polarized lepton collider or 
a neutrino factory.

The MOLLER design shares many similarities with E158 and Qweak. 
An 11 GeV electron beam in JLab Hall A will be incident on a 1.5 meter LH$_2$ target. 
A toroidal spectrometer would exploit the unique topology of M\o ller scattering involving 
identical particles, avoiding the typical 50\%\ azimuthal acceptance loss associated with coil 
placement, by employing an odd number of coils and collecting scattered electrons 
from both the forward and backward directions in the center-of-mass frame. 
The M\o ller-scattered electrons in the full range of the azimuth would be directed to a ring 
focus 30 m downstream of the target. 
The detector system incorporates a great deal of redundancy to monitor the principal 
backgrounds from electron-proton elastic and inelastic scattering to better than 1\% accuracy. 
The prediction for $A_{PV}$ is 35 ppb and the aim is to achieve a statistical uncertainty 
of 2.1\%. 

MOLLER will greatly benefit from the steady improvement in the techniques employed to measure 
parity-violating asymmetries to sub-ppb systematic precision and to also achieve normalization 
control at the sub-\%\ level. 
For example, two redundant monitors (one continuous) of electron beam polarization 
will be employed to achieve 0.4\% fractional accuracy. 
Auxiliary detectors would track individual particles at low rates to measure 
$\langle Q^2\rangle$ to 0.5\% fractional accuracy. 
Very forward angle detectors downstream of the main detectors will verify that luminosity 
fluctuations due to jitter in electron beam properties and target density are under control. 
All three methods to reverse the sign of the asymmetry that have been developed for previous 
experiments ($g-2$ spin flip, half-wave plate insertion, and the double Wien filter) will also 
be employed periodically.
The construction project is a US Department of Energy Major Item of Equipment (MIE), with 
additional funding from the Canada Foundation for Innovation and the US National Science 
Foundation. 
The engineering design is complete, along with the acquisition of the majority of fabricated 
equipment. 
Installation has begun in late 2025 and first physics is expected in Spring 2027.

\subsubsection{P2}
The P2 experiment is an approved project for the newly funded MESA facility at 
Mainz~\cite{Becker:2018ggl}. 
The goal is $\delta(A_{PV}) = \pm 1.7\%$ (total experimental) for elastic electron-proton 
scattering, which would yield $\delta(Q_W^p)\simeq 2\%$ and 
$\delta(\sin^2\theta_W) \simeq \pm 0.15\%$. 
To achieve the statistics would require a 155 MeV, 150 $\mu$A beam incident on 
a 60 cm LH$_2$ target for 10,000 hours.
Apart from the improvement in statistical reach and hence sensitivity to new physics over 
the JLab Qweak experiment, the lower beam energy reduces theoretical uncertainties in extracting 
$\sin^2\theta_W$. 
P2 will strengthen the constraints \eqref{Qweak} to reach 
$\Lambda \sim {\cal O} (50 \mbox{ TeV})$.

The design requires a solenoidal magnet downstream of the target which would focus scattered 
electrons within $10^\circ<\theta_{\rm lab}<30^\circ$ in the full range of the azimuth onto 
integrating Cherenkov detectors.  
The field would sweep out the large M\o ller electron background and an inner shield protects 
the detectors from the target's direct photon background. 
The theoretical prediction is $A_{PV}\sim 20$ ppb and 
$\delta_{\rm stat}(A_{PV})$ is $\pm 0.25$ ppb. 
The total rate in the detectors would approach 0.5~THz. 

The design must overcome many technical challenges such as controlling electron beam 
fluctuations at the sub-nm level and controlling target density fluctuations 
to a few parts in $10^{-5}$. 
A new method to measure the electron beam polarization could be potentially employed 
which would require a novel polarized hydrogen gas target~\cite{Chudakov:2004de}.
The project is currently under construction at the MESA facility. 
Commissioning of the experiment is anticipated to begin in 2027.

\subsection{Future Prospects}
\subsubsection{SoLID}
The SoLID experiment~\cite{Chen:2014psa} at JLab has been proposed to make a series of $A_{PV}$ 
measurements with 0.5--1\%\ fractional accuracy in deep inelastic scattering of electrons 
off $^2$H. 
The primary motivation is to measure the poorly known neutral current axial-vector quark 
couplings $g_{VA}^{eq}$, resulting in access to complementary TeV-scale lepton-quark 
interactions. 
It would also result in a measurement of weak mixing angle with an uncertainty of 
$\delta(\sin^2\theta_W) \approx 0.0006$ at $Q\sim 2.5$ GeV.

The heart of the apparatus is a large acceptance solenoid which facilitates $A_{PV}$ 
measurements in narrow Bjorken-$x$ bins ($0.3 \lesssim x \lesssim 0.7$) with a lever arm of 
a factor of 2 in $Q^2$ while keeping $W^2 > 4$ GeV$^2$ and $\langle Q^2\rangle\sim 5$ GeV$^2$. 
The $\mathrm{LD}_2$ target would be placed inside the solenoidal field and several planes 
of absorbing material between the target and detectors with slits would tailor 
the momentum acceptance to the kinematic region of interest.  

One important feature is that, unlike other $A_{PV}$ measurements that integrate detector 
signals over different helicity periods, high precision hit-based tracking with gas electron 
multiplier detectors is required to reconstruct the scattering angle and momentum of scattered 
electrons.  
To separate electrons from background, predominantly a 100 times more $\pi^-$ particles, 
particle identification would be performed with heavy-gas Cherenkov detectors placed 
symmetrically about the beam axis inside the solenoid. 
An electromagnetic calorimeter would provide the primary electron trigger as well as allow 
additional pion rejection. 

This experiment was approved by the JLab Program Advisory Committee and a significant amount 
of R\&D has been carried out to flesh out a complete technical design of the apparatus. 
The project has been strongly endorsed by the nuclear physics community and is highlighted as 
one of the projects that would go forward if new funding becomes available at the federal level 
for future projects. 
Constraints on new physics scales $\Lambda \sim 22$~TeV can be expected from SoLID which will 
target the parity-flipped combination $2 g_{VA}^{eu} - g_{VA}^{ed}$.

\subsubsection{MESA Carbon}
The P2 detector can also be used to measure the asymmetry in electron scattering on $^{12}$C.
It is complementary to $ep$ scattering both regarding experimental systematics and 
phenomenological implications.
Indeed, while the $ep$ asymmetry benefits from the $1 - 4 \sin^2\hat\theta_W$ suppression, both 
cross section and asymmetry are tiny and two independent nucleon form factors 
--- the Dirac form factor $F_1$ and the Pauli form factor $F_2$,
or equivalently the electric and magnetic form factors $G_E$ and $G_M$ --- are needed as input.
On the other hand, the isoscalar nature of $^{12}$C implies that the asymmetry is (at lowest 
order) proportional to $\sin^2\hat\theta_W$ itself, but rates and $A_{PV}$ are much larger.
Furthermore, since the $^{12}$C ground state is also a Lorentz scalar, only one form factor 
enters~\cite{Souder:1990ia} which may be constrained~\cite{Koshchii:2020qkr,Cadeddu:2024baq} 
experimentally by means of an additional kinematic setting (by backward scattering).
The challenge here consists of the necessity of an exquisite polarization measurement to 
a precision of a few permille compared to the typical percent level polarimetry for the $ep$ and 
$ee$ asymmetries, but as an upside, there would only be a very small residual experimental 
correlation between these measurements.

Thus, the $^{12}$C asymmetry would provide an interesting alternative approach to measure 
$\sin^2\hat\theta_W$ (possibly more precisely than E158 or Qweak), but its most important impact 
would be its sensitivity to an independent linear combination of effective couplings, 
$g_{AV}^{eu} + g_{AV}^{ed}$, with an associated sensitivity to new physics very similar to 
the hydrogen case. 
This linear combination is not very different from the one extracted from atomic parity 
violation (see Sec.~\ref{APV} below), but the latter is exposed to an intricate many-body QED 
bound state calculation and the associated hard-to-estimate theory uncertainty.

\subsubsection{The Electron Ion Collider}
The Electron Ion Collider (EIC)~\cite{Accardi:2012qut, AbdulKhalek:2021gbh} is under 
construction at Brookhaven National Laboratory with the primary goal to gain detailed 
insights into the internal structure of the nucleon such as how the proton mass and spin emerge 
from the dynamics of quarks and gluons~\cite{Deshpande:2005wd}. 
A detailed detector design has been conceived to study the deep-inelastic scattering process 
when a 10 to 18~GeV polarized electron beam is brought into collision with 100 to 250 GeV 
polarized nucleons, providing a center of mass energy up to 140 GeV.
With the envisioned detector capabilities, weak neutral current interactions over a large range 
of $Q^2$ will be measured. 

Nucleon structure in inelastic electron-nucleon neutral current scattering is characterized by 
structure functions that are functions of $Q^2$ and $W^2$, which are inferred from the initial 
and final state electron 
4-momenta~\cite{Anselmino:1993tc, Anselmino:1994gn, Ji:1993ey, Stratmann:1995fn}. 
In the deep inelastic regime ($Q^2\gg 1$ GeV$^2$ and $W^2\gg 4$~GeV$^2$), one can invoke 
the quark-parton model where these structure functions can be analyzed to extract not only the quark probability distribution 
functions (PDFs) and but also the fundamental electroweak couplings. 
Detailed studies have been carried out of how well the structure functions can be measured once 
data samples in excess of 100 fb$^{-1}$ become available~\cite{Zhao:2016rfu, Boughezal:2022pmb}. 
In particular, proton and neutron structure functions at moderately high Bjorken-$x$ can be used 
to extract u- and d-quark PDFs and extract measurements of the weak mixing angle 
in an entirely new $Q^2$ regime in the range $50<Q<80$ GeV, just below the Z-resonance. 
Apart from confirming the running of the weak mixing angle, such measurements have 
the potential, as studied in a SMEFT analysis~\cite{Boughezal:2022pmb}, of accessing unique 
lepton-quark dimension six operators that can lift flat directions in a global analysis 
of Drell-Yan processes at the LHC.

\subsubsection{SuperKEKB Collider}
A new program dubbed ``Chiral Belle" has been proposed to enhance the physics program of 
the Belle-II detector and upgraded high luminosity of the KEK B-Factory electron positron 
collider~\cite{USBelleIIGroup:2022qro, Roney:2025lwo}. 
With the addition of three new hardware components, namely a polarized electron source, a set of 
spin rotator magnets and a Compton polarimeter to monitor the electron beam polarization, 
precision measurements of weak neutral current coupling combinations become feasible 
via left-right asymmetries in the process $\vec{e}^{\, -} e^+\rightarrow f^+f^-$. 
The detector can cleanly identify pure samples to the three leptonic final states, 
as well as to $c$- and $b$-quark final states. 

The neutral current vector couplings for $e$ and $\tau$ would be measured to similar precision 
as that at the $Z$ pole, while those for $\mu$, $b$ and $c$ would be vastly improved. 
In particular, since these measurements would be made at the $\sim 10$ GeV scale, the running of 
the weak mixing angle will be tested for a third generation fermion for the first time. 

In addition, via the electron weak neutral current vector coupling, there would then be 
a total of three measurements of the weak mixing angle, namely at 100 MeV (MOLLER), 
10 GeV (SuperKEKB), and 90 GeV ($Z$ pole), each with $\sim 0.1$\%\ precision, providing 
great sensitivity to a variety of new physics scenarios at the MeV and TeV scales.

\section{Atomic Parity Violation}
\label{APV}
Another way to access electroweak and possibly new physics~\cite{Marciano:1990dp} 
is to investigate parity-violating asymmetries $A_{\rm PV}$ in atoms.
For example, permanent electric dipole moments are clear signals of parity violation (PV).
However, they require in addition the violation of time reversal symmetry and 
have not been observed, so far.
Thus, one focuses on electric dipole {\em transition\/} amplitudes $E_1$.
These are strictly forbidden by the parity-conserving electromagnetic interaction, 
and the leading contribution is due to $Z$ boson exchange.
The $E_1^{\rm PV}$ amplitudes interfere with the allowed magnetic dipole transition amplitudes 
$M_1$~\cite{Guena:2005uj}, generally resulting in minute values of $A_{\rm PV}$ for light atoms. 
As a reference, for deuterium one has
\begin{equation}
    A_{\rm PV} \approx 2 \real \left[ \frac{E_1^{\rm PV}}{M_1} \right] 
    \sim \frac{\alpha m_e^2}{2 \pi v^2} \approx 5 \times 10^{-15}
\label{deuteronAPV}
\end{equation}

On the other hand, in heavy atoms the parity-violating transition rates grow faster than with 
the third power of the number of protons ($Z$) and are strongly enhanced~\cite{Bouchiat:1974kt}.
Experiments searching for atomic parity violation (APV) were performed with an optical rotation 
approach on thallium ($Z = 81$)~\cite{Edwards:1995zz,Vetter:1995vf}, 
lead ($Z = 82$)~\cite{Meekhof:1993zz}, and bismuth ($Z = 83$)~\cite{Macpherson:1991opp}, 
to extract the so-called nuclear weak charge, $Q_W(Z,N) = (1 - 4 \sin^2\theta_W) Z - N$ 
(given in terms of $Z$ and the neutron number $N$), which is modified by the one-loop 
electroweak radiative corrections given in Ref.~\cite{Marciano:1982mm}. 
However, for precision tests one also needs accurate calculations in many-body atomic 
physics~\cite{Ginges:2003qt,Derevianko:2006sf}, and these elements possess a complicated atomic 
structure. 

Another way to enhance $A_{\rm PV}$ in Eq.~\eqref{deuteronAPV} is to consider highly suppressed 
$M_1$ amplitudes.
For example, the two radial wave functions involved in the $7S_{1/2}$ to $6S_{1/2}$ transition 
in cesium ($Z = 55$) or the $7P_{1/2}$ to $6P_{1/2}$ transition in thallium are orthogonal, 
so that $M_1$ cancels up to small relativistic effects~\cite{Bouchiat:1974kt}, 
lifting $A_{\rm PV}$ to the $10^{-6}$ level.
In the case of cesium one then magnifies the interference of $E_1^{\rm PV}$ by applying 
an external electric field $\vec E$.
As a consequence, the Stark-induced electric dipole amplitude, $E_1^{\rm ind} = \beta \vec E$, 
where $\beta$ is the tensor transition polarizability, 
now plays the r$\hat{\rm o}$le of the $M_1$ amplitude in the asymmetry \eqref{deuteronAPV}.
Notice that the sign of $E_1^{\rm ind}$ can be flipped by reversing the orientation of~$\vec E$.
A complication of the Stark-interference technique is that besides a measurement of $A_{\rm PV}$ 
and a calculation of $E_1$, a precise determination of $\beta$ is required, as well.
Nevertheless, this whole program has been carried out successfully for the case 
of $^{133}_{\phantom0 55}{\rm Cs}$, where there are currently two semi-empirical approaches 
to $\beta$ of similar precision.
The ratio of the off-diagonal hyperfine amplitude $M_1^{\rm hfs}$ to the tensor transition 
polarizability has been measured by the Boulder group~\cite{Bennett:1999pd}.
Combined with the relativistic many-body calculation~\cite{Dzuba:2000gf} of $M_1^{\rm hfs}$ 
one obtains $\beta = (26.957 \pm 0.044_{\rm\, exp} \pm 0.027_{\rm\, th})\, a_0^3$, 
where $a_0$ is the Bohr radius.
Alternatively, one combines~\cite{Dzuba:1997df} the measurement of the ratio of scalar to 
tensor transition polarizabilities~\cite{Cho:1997zz} with the recent calculation~\cite{TranTan:2023kll} 
of the scalar polarizability to obtain 
$\beta = (26.887 \pm 0.030_{\rm\, exp} \pm 0.021_{\rm\, th})\, a_0^3$.
The two determinations average to 
$\beta = (26.911 \pm 0.025_{\rm\, exp} \pm 0.017_{\rm\, th})\, a_0^3$.
One can then write,
\begin{equation}
	Q_W (55,78) = N \left( \frac{{\rm Im}\, E_1^{\rm PV}}{\beta} \right)_{\rm\, exp} 
    \left( \frac{|e|\, a_0}{{\rm Im}\, E_1^{\rm PV}} \frac{Q_W}{N} \right)_{\rm th} 
	\left( \frac{\beta}{a_0^3} \right)_{\rm\, exp + th} \left( \frac{a_0^2}{|e|} \right)
\end{equation}
The uncertainties associated with the atomic wave function calculations are also relatively 
small for cesium~\cite{Roberts:2014bka,Safronova:2017xyt}, where a state-of-the-art many-body 
atomic structure computation resulted in 
${\rm Im}\, E_1^{\rm PV}=(0.8977\pm 0.0040)\times 10^{-11}|e|\, a_0\, Q_W/N$~\cite{Dzuba:2012kx}.
The experiments~\cite{Wood:1997zq,Guena:2004sq} computed a weighted average of the results 
of two $\Delta F = \pm 1$ hyperfine lines to cancel the nuclear spin-dependent 
contribution\footnote{Ref.~\cite{Wood:1997zq} also found a non-zero difference between 
the two hyperfine transitions with 7$\sigma$ significance.  
It is expected to arise mainly due to the nuclear anapole moment~\cite{Haxton:2001ay}.}.
The combination, ${\rm Im}\, E_1^{\rm PV}/\beta = - 1.5924 \pm 0.0055$~mV/cm,
is dominated by the more precise result of Ref.~\cite{Wood:1997zq}, and implies,
\begin{equation}
	Q_W (55,78)=- 72.41\pm 0.26_{\rm\, exp} \pm 0.33_{\rm\, th}\ = - 72.41 \pm 0.42_{\rm\, total}
\label{QWCs}
\end{equation}
or equivalently the constraint, $55 g_{AV}^{ep} + 78 g_{AV}^{en} = 36.25 \pm 0.21$, which is 
2.0 standard deviations below the SM prediction $Q_W (55,78) = - 73.26 \pm 0.01$~\cite{PDG10}.
Within the SM this can also be translated into a determination of the weak mixing angle, 
$\hat s^{\, 2} (2.4\mbox{ MeV}) = 0.2349 \pm 0.0018$, where the scale setting 
$\mu \sim 2.4\mbox{ MeV}$ is based on an estimate~\cite{Bouchiat:1983uf} for the typical 
momentum transfer for parity violation experiments in Cs (the corresponding estimate 
for Tl amounts to 8~MeV).
Since the theoretical atomic structure uncertainties are 3\% for thallium~\cite{Dzuba:1987px} 
and even larger for the other atoms~\cite{Dzuba:2001zz}, the 0.6\% overall determination 
in~\eqref{QWCs} is currently the only precision determination of the weak charge of any nucleus 
containing neutrons.
However, two atomic structure calculations by independent groups 
found results that are in tension with the above, reporting 
${\rm Im}\, E_1^{\rm PV}=(0.8906\pm 0.0024)\times 10^{-11} |e|\, a_0\, 
Q_W/N$~\cite{Porsev:2009pr} and ${\rm Im}\, E_1^{\rm PV} = 
(0.8893 \pm 0.0027) \times 10^{-11} |e|\, a_0\, Q_W/N$~\cite{Sahoo:2021thl} instead,
and implying a value of $Q_W (55,78)$ in good agreement with each other and with the SM.
The disagreements triggered an exchange between two of the
groups~\cite{Roberts:2021esp,Sahoo:2022qdu}, as well as a further 
publication~\cite{Chakraborty:2023}.
This illustrates the great challenges associated with the atomic 
structure calculations, but there are possible prospects to reduce 
the uncertainty in $E_1^{\rm PV}$ to the level of 0.2\% by 
a parity-mixed coupled-cluster approach~\cite{Tan:2021gav}.

A feasible alternative to cesium appears to be francium.
From the experimental side, there has been recent progress in the pursuit of a program in 
the analogous $8S$ to $7S$ transition, {\em i.e.}, to take sums and differences of 
the two hyperfine transitions to extract the weak charge and the anapole moment, respectively.
It is conceivable that a measurement with per-mille statistical uncertainty can be obtained in 
the coming years (the systematics is under investigation).
On the theory side, existing codes for Cs can be adapted for the case of Fr.
However, the atomic structure calculations are expected to come with larger error bars 
than for Cs.
This is because Fr has an additional 32 electrons and it contains F-shell electrons 
while Cs does not.
Another complication may be a lower quality of supporting data, such as various matrix elements, 
entering a future Fr calculation, as Cs has been studied extensively 
in part because of Cs atomic clocks. 
Indeed, the basis for the quantification of theory errors are typically not only 
self-consistency and convergence rates, but also the agreement with experimental data 
for supporting quantities.
Finally, there is an enhanced sensitivity of Fr to the neutron skin which scales as $Z^2$ 
and has been estimated to 0.6\%~\cite{Brown:2008ib}.
Thus, APV in Fr may alternatively be viewed as a determination of neutron skins.

Interestingly, the atomic structure uncertainties are known to mostly cancel if one takes ratios 
of the parity violating effects in different isotopes~\cite{Rosner:1995aj}. 
The first result of this type of experiment was announced recently by 
the Mainz group~\cite{Antypas:2018mxf}, who studied APV in 
$^{170}$Yb, $^{172}$Yb, $^{174}$Yb, and $^{176}$Yb, at the 0.5\% level.
The resulting three ratios can be interpreted as a determination of 
$\hat s^{\, 2} = 0.258 \pm 0.052$, and represent a very complementary approach 
to search for BSM physics~\cite{Ramsey-Musolf:1999qyv}.
If the precision increases in the future, one would ultimately face uncertainties from 
differences in the neutron charge radii~\cite{Pollock:1992mv,Chen:1993fw}. 
These can be constrained experimentally~\cite{Horowitz:1999fk}, {\em e.g.}, by measuring 
parity-violating elastic electron scattering asymmetries off heavy nuclei as done at JLab by the PREX collaboration
on $^{208}$Pb~\cite{PREX:2021umo} and the CREX collaboration on $^{48}$Ca~\cite{CREX:2022kgg}.

Another possible avenue to take in the future is to measure APV on a single trapped ion.
The proof-of-principle of this technique has already been demonstrated 
with Ba$^+$ ions~\cite{Sherman:2005}.
The $7S_{1/2}$ to $6D_{3/2}$ transition amplitude in Ra$^+$ ($Z = 88$) appears particularly 
promising, as the required lasers are all at convenient wavelengths~\cite{Wansbeek:2008yma}.
More importantly, while in $S$ to $S$ transitions like in cesium there are significant 
cancellations between initial and final states, the parity-violating mixing in Ra$^+$ is 
strongly dominated by a single intermediate state so that one expects more stable results from 
the atomic structure calculations.
This also increases the overall APV signal which in total is about a factor of 50 stronger 
in Ra$^+$ compared to Cs.
Taken together, it is not inconceivably that the overall uncertainty in the weak charge 
of $^{226}$Ra can be controlled at the per mille level.
At this level of precision, the charge radius of $^{226}$Ra would be needed to good accuracy.
Fortunately, it has been demonstrated that charge radii of radioactive isotopes can be measured 
precisely using muonic atoms~\cite{Skawran:2019krq}.

\section{Neutrino Scattering}
\label{neutrinos}
\subsection{Conventional Neutrino Scattering}
Neutrino scattering played a key role in the establishment of the SM.
Besides discovering neutral currents, the Gargamelle heavy liquid bubble 
chamber~\cite{GargamelleNeutrino:1973jyy} provided the first (even if crude) determination 
of $\sin^2\theta_W$ from the ratio of cross sections of neutral and 
charged current $\nu_\mu$ interactions.
The neutrino beam consisted of $\nu_\mu$ and $\bar\nu_\mu$ with energies $E_\nu$ 
in the single GeV range.

The first precise results --- initially at the 10\% level --- were obtained 
in neutrino-nucleus DIS ($\nu$-DIS) by the CDHS~\cite{Blondel:1989ev} 
and CHARM~\cite{CHARM:1987pwr} Collaborations to about 2\%.
Neutrino beams with $E_\nu$ up to 200~GeV were directed at approximately isoscalar targets.
The CCFR~\cite{CCFR:1997zzq} Collaboration at Fermilab achieved similar results with beam 
energies up to 600~GeV.
It morphed later into the NuTeV Collaboration~\cite{NuTeV:2001whx} which had an almost pure 
$\nu$ ($\bar\nu$) beam available with only a small contamination 
of 0.03\% $\bar\nu$ (0.4\% $\nu$) and a moderate $\nu_e$ content of about 1.7\%.
These exquisite conditions allowed to determine the cross section ratios $R_\nu$ and 
$R_{\bar\nu}$ individually, and to quote the result 
(for the on-shell definition $\cos\theta_W \equiv M_W/M_Z$),
\begin{equation}
    \sin^2\theta_W = 0.2277 \pm 0.0013_{\rm stat} \pm 0.0009_{\rm syst}
\label{nutev}
\end{equation}
This represents a $2.7~\sigma$ conflict, however, with the SM prediction of 
$\sin^2\theta_W = 0.22348 \pm 0.00010$.
Following this, several experimental and theoretical developments suggested significant 
modifications of the extracted value, with most of them reducing the discrepancy.
\begin{description}
\item[PDFs] NuTeV also reported on a measurement~\cite{NuTeV:2007uwm} of the difference $S^-$ 
between the strange and antistrange quark momentum distributions (from di-muon events) which 
differed in sign from an earlier result~\cite{NuTeV:2002ryj}, and in size from 
theoretical considerations~\cite{Bentz:2009yy} (favoring a much smaller effect).
The result~\cite{NuTeV:2007uwm} translated into a shift~\cite{NuTeV:2002ryj} 
$\delta \sin^2\theta_W = - 0.0014 \pm 0.0010$. 

Parton distribution functions violate isospin symmetry potentially at levels much stronger 
than expected~\cite{Sather:1991je,Rodionov:1994cg}, but are currently not well constrained 
phenomenologically and within uncertainties the NuTeV anomaly could be accounted for in full 
or conversely made even larger~\cite{Martin:2003sk}.
On the other hand, the leading contribution from quark mass diﬀerences turned out 
to be approximately model-independent~\cite{Londergan:2003ij} and would correspond to a shift 
almost identical to the one from $S^-$.
QED splitting eﬀects also violate isospin symmetry and model-independently reduce the value 
in Eq.~\eqref{nutev}.
A calculation~\cite{Gluck:2005xh} suggested to reduce it by 
$\delta \sin^2\theta_W = - 0.0011$, but the associated uncertainty is difficult to estimate.

\item[Nuclear effects] 
Nuclear shadowing~\cite{Kopeliovich:2012kw} is likely to aﬀect the interpretation of the result 
somewhat, but NuTeV argued that their data are dominated by $Q^2$ values at which it is 
expected to be small.
However, another nuclear eﬀect, the isovector EMC eﬀect~\cite{Cloet:2009qs}, is considerably 
larger as it aﬀects all neutrons in the nucleus and not just excess neutrons.
It would also model-independently reduce the discrepancy and has been estimated to induce 
a shift~\cite{Bentz:2009yy} in Eq.~\eqref{nutev} by $\delta\sin^2\theta_W = - 0.0019 \pm 0.0006$.

\item[Radiative corrections] $\sin^2\theta_W$ may also shift at the level of the quoted 
uncertainty when analyzed using QED and electroweak radiative 
corrections~\cite{Arbuzov:2004zr,Diener:2005me}, as well as QCD corrections to the structure 
functions~\cite{Dobrescu:2003ta} that became available later.
However, their precise impact can be estimated only by a simultaneous fit of the NuTeV data with 
a new set of PDFs including these radiative corrections, isospin breaking effects, 
and an asymmetric strange seas. 

Missing radiative corrections have been estimated~\cite{Diener:2005me} to induce uncertainties
in the extracted $\sin^2\theta_W$ of about $\pm 0.0005$, much larger than assumed initially. 

\item[$\nu_e$ beam content] The measured branching ratio for $K \longrightarrow \pi e \nu_e$ 
decays enters crucially in the determination of the $\nu_e~(\bar\nu_e)$ contamination of 
the $\nu_\mu~(\bar\nu_\mu)$ beam, but moved from $4.82 \pm 0.06\%$ at the time of 
Ref.~\cite{NuTeV:2001whx} to the current value of $5.07 \pm 0.04\%$, {\em i.e.}, 
a change by more than 4~$\sigma$. 
This implied an estimated shift similar to the one from $S^-$, but in the opposite direction.
\end{description}
These considerations show that the entire suite of $\nu$-DIS experiments 
--- while impressive --- are not competitive as high-precision determinations 
of $\sin^2\theta_W$ at least until a re-analysis using PDFs including updated experimental and 
improved theoretical information and correlations has been completed.
These data may continue to be of relevance for the study of $\nu$-oscillations, 
though~\cite{Ansari:2021cao}.

Free of all these complications is the scattering of neutrinos from charged leptons.
The downside here are the much smaller cross sections compared to heavy nuclei (typically iron) 
discussed above.
The total cross-section for $\nu_\mu e \to \nu_\mu e$ or $\bar\nu_\mu e \to \bar\nu_\mu e$ 
elastic scattering (for $E_\nu \gg m_e$) is~\cite{Formaggio:2012cpf},
\begin{equation}
	\sigma = \frac{G^2_F m_e E_\nu}{2 \pi} \left[ (g^{\nu e}_{LV} \pm g^{\nu e}_{LA})^2 + 
    \frac{1}{3} (g^{\nu e}_{LV} \mp g^{\nu e}_{LA})^2 \right]
\label{nuXsec}
\end{equation}
where the upper and lower signs refer to $\nu_\mu$ and ${\bar\nu}_\mu$, respectively.
The most accurate measurements of $\sin^2 \theta_W$ from $\nu$-lepton scattering are from  
$R\equiv\sigma_{\nu_{\mu} e}/\sigma\/\/_{\bar\nu_{\mu} e}$, where many of the systematic 
uncertainties cancel. 
The results are 
$\sin^2 \theta_W = 0.211 \pm 0.037$~\cite{CHARM:1988tlj} and
$\sin^2 \theta_W = 0.2324 \pm 0.0083$~\cite{CHARM-II:1994dzw} (CERN), and
$\sin^2 \theta_W = 0.195 \pm 0.022$~\cite{Ahrens:1990fp} (BNL),
where radiative corrections (other than effects quadratic in the top quark mass) 
are small compared to the experimental precision and negligible. 
From the data one can also determine $g^{\nu e}_{LV,LA}$ and set limits (even though weak) 
on new physics.

The cross-sections for $\nu_e$-$e$ and $\bar\nu_e$-$e$ are obtained from Eq.~\eqref{nuXsec} by 
replacing $g_{LV,LA}^{\nu e}$ by $g_{LV,LA}^{\nu e} + 1$, where the +1 is due to 
the charged-current contribution.
The $\bar\nu_e$-$e$ elastic scattering cross section has been measured by 
the TEXONO Collaboration~\cite{TEXONO:2009knm} using reactor neutrinos originating from 
the Kuosheng Nuclear Power Plant, with the result 
$\sin^2 \theta_W = 0.251 \pm 0.039$ at the very low scale $\mu$ of about 2~MeV.

There are prospects~\cite{deGouvea:2019wav} to measure $\sin^2 \theta_W$ in $\nu_e$-$e$ 
scattering with the near detector at DUNE.
A precision of about 2\% may be possible at an average scale of $\mu = 55$~MeV.
Additional information may be provided by neutrino-trident scattering, 
$\nu A \to \nu A \ell^+ \ell^-$, and also by the movable near detector DUNE--PRISM.

\subsection{Unconventional Neutrino Scattering}
\subsubsection{Coherent Elastic Neutrino Nucleus Scattering}
Soon after the discovery of weak neutral currents in neutrino 
scattering~\cite{GargamelleNeutrino:1973jyy}, it was pointed out that a neutrino could scatter 
coherently off an entire nucleus so long as the wavelength associated with the momentum 
transferred was significantly larger than the nuclear size~\cite{Freedman:1973yd}, nowadays 
called coherent elastic neutrino-nucleus scattering or CE$\nu$NS.  
The development of techniques over the last couple of decades to observe few keV nuclear recoils 
to search for cold dark matter made feasible the first observation of this process by 
the COHERENT experiment at the Spallation Neutron Source (SNS) using 
a sodium-doped cesium-iodide detector~\cite{COHERENT:2017ipa}.  

Dedicated reactor and stopped pion beam experiments as well as dark matter experiments are now 
actively making measurements of the CE$\nu$NS process on a variety of nuclei.
While such measurements are of primary interest to look for non-standard neutrino interactions, 
precise measurements of the cross-section on a variety of nuclei have implications for 
the neutron distribution of nuclei, searches for MeV scale dark bosons, and solar and other 
astrophysics neutrino fluxes, especially when combined with other measurements of the nuclear 
weak form factor using electron scattering and measurements of the weak charge of nuclei in APV 
(Sec~\ref{APV} above)~\cite{Abdullah:2022zue,AtzoriCorona:2024vhj}. 
While these measurements are currently not competitive for measurements of the weak mixing 
angle~\cite{Maity:2024aji} and precision tests of neutral current interactions, measurements 
at future large-scale dark matter detectors might well become relevant especially if there 
remain unexplained anomalies in measurements of nuclear neutron distributions, 
other low-energy measurements of the weak mixing angle or APV~\cite{AtzoriCorona:2025ibl}.

\subsubsection{Collider Neutrinos}
It is also possible to study the neutrinos (and other long-lived weakly interacting particles) 
produced at the collision point at the ATLAS detector at the LHC, by placing various detectors 
downstream in the line of sight of one of the proton beams.
Such neutrinos with energies up to several TeV have already been observed in charged current 
interactions by the FASER~\cite{FASER:2023zcr} and SND@LHC~\cite{SNDLHC:2023pun,SNDLHC:2024qqb} 
Collaborations.
It is possible that experiments at a Forward Physics Facility (FPF) at CERN will be able
to also detect neutral current $\nu$-interactions.
In fact, it was estimated in a recent study~\cite{MammenAbraham:2023psg} that $\sin^2 \theta_W$ 
can conceivably be measured at $\mu\sim 10$~GeV with an uncertainty of about 3\% provided 
neutrino-nucleus cross section uncertainties can be kept sufficiently small.

\section{Summary and Outlook}
Measurements of the weak mixing angle played a key role in the establishment of the Standard 
Model, and high-precision measurements of this parameter today constitute one of the most 
sensitive probes of physics beyond it.
As a derived quantity, the experimental results can be directly compared with the SM prediction,
which is particularly interesting as its value shifts in most models of new physics, especially 
those addressing the hierarchy problem 
(the tiny value of the Higgs mass compared to the fundamental Planck scale).

In the presence of new physics, the patterns of deviations of the values of the weak mixing 
angle extracted from different observables may hint at the underlying new physics model.
This is where the complementarity of the various types of experiments discussed here 
reveals its full potential.  
In particular, the parity-violating electron scattering experiments that will start to take data 
relatively soon will reach precisions comparable to those achieved by the $Z$ pole experiments 
at the LEP and SLC colliders. 
These are the MOLLER ($ee$) and P2 ($ep$ and $e$C elastic)
fixed-target experiments under construction or planned at the CEBAF (JLab) and MESA (Mainz) 
accelerators.
There are also proposals awaiting funding for SoLID to measure parity-violating asymmetries in $ed$ deep-inelastic scattering 
at JLab as well as for very high-precision measurements at the KEKB collider with 
a polarized $e^-$ beam in Japan.
New ideas and R\&D continues in the study of parity violation in atoms and ions with the longer term goal of providing measurements complementary to the pioneering Cesium result.

Conventional neutrino deep-inelastic scattering experiments currently suffer from 
an insufficient theoretical understanding of heavy nuclei, limiting its usefulness in 
electroweak precision physics.
On the other hand, it became possible recently to observe and make quantitative measurements in 
coherent elastic neutrino nucleus scattering, as well as to detect the neutrinos produced 
at very high energies at the hadron collider LHC.
While the precision of future extractions of the weak mixing angle are significantly less precise than other observables discussed above, they would provide valuable complementary insights.

\section*{Acknowledgements}
This project has received funding from the European Research Council (ERC) under the European 
Union’s Horizon Europe research and innovation programme (Grant agreement No. 101142600),
by the Deutsche Forschungsgemeinschaft (DFG, German Research Foundation) --- 
project number: 514321794 (CRC1660: Hadron and Nuclei as discovery tools), 
and by the US Department of Energy Award DE-SC0021035.
JE is also supported by the Cluster of Excellence 
{\em Precision Physics, Fundamental Interactions, and Structure of Matter\/} 
(PRISMA$^{++}$ EXC 2118/2) funded by the German Research Foundation (DFG) 
within the German Excellence Strategy (Project ID 390831469).

\end{document}